\documentclass[10pt,conference]{IEEEtran}
\IEEEoverridecommandlockouts

\usepackage{cite}
\usepackage{amsmath,amssymb,amsfonts}
\usepackage{graphicx}
\usepackage{textcomp}
\usepackage{xcolor}
\def\BibTeX{{\rm B\kern-.05em{\sc i\kern-.025em b}\kern-.08em
    T\kern-.1667em\lower.7ex\hbox{E}\kern-.125emX}}

\usepackage{soul}
\usepackage{hyperref}
\usepackage{multirow}
\usepackage{fancybox}
\usepackage{enumitem}
\usepackage{balance}
\usepackage{booktabs}
\usepackage{color}
\usepackage{float}
\usepackage{xcolor}
\usepackage{listings}
\usepackage{tabularx}
\usepackage{adjustbox}
\usepackage{array}
\usepackage{amsmath}
\usepackage{xspace}
\usepackage{url}
\usepackage{tikz}
\usepackage{caption}
\usepackage{listings}
\usepackage{color}
\usepackage{graphicx}
\definecolor{dkgreen}{rgb}{0,0.6,0}
\definecolor{gray}{rgb}{0.5,0.5,0.5}
\definecolor{mauve}{rgb}{0.58,0,0.82}
\usepackage{pgf-pie}
\usepackage{algorithm}
\usepackage{algpseudocode}
\usepackage{subcaption}
\usepackage[normalem]{ulem}

\usepackage{tcolorbox}

\lstdefinestyle{CppStyle}{
  language=C++,
  basicstyle=\ttfamily\footnotesize,
  keywordstyle=\color{blue},
  commentstyle=\color{green!50!black},
  stringstyle=\color{red},
  numbers=left,
  numberstyle=\tiny\color{gray},
  stepnumber=1,
  breaklines=true,
  tabsize=2,
  showspaces=false,
  showstringspaces=false
}

\lstdefinestyle{PythonStyle}{
  language=Python,
  basicstyle=\ttfamily\footnotesize,
  keywordstyle=\color{blue}\bfseries,
  commentstyle=\color{green!50!black},
  stringstyle=\color{red},
  numbers=left,
  numberstyle=\tiny\color{gray},
  stepnumber=1,
  breaklines=true,
  tabsize=4,
  showspaces=false,
  showstringspaces=false
}

\makeatletter
\newcommand{\mybox}[1]{%
	\setbox0=\hbox{#1}%
	\setlength{\@tempdima}{\dimexpr\wd0+13pt}%
	\begin{tcolorbox}[boxrule=0.5pt, colback=white, arc=4pt,
		left=6pt,right=6pt,top=6pt,bottom=6pt,boxsep=0pt]
		#1
	\end{tcolorbox}
}
\usepackage[framemethod=default]{mdframed}
\definecolor{lightblue}{RGB}{227,242,253}
\definecolor{codegreen}{rgb}{0,0.6,0}
\definecolor{codegray}{rgb}{0.5,0.5,0.5}
\definecolor{codepurple}{rgb}{0.58,0,0.82}
\definecolor{backcolour}{rgb}{0.95,0.95,0.92}

\lstdefinestyle{mystyle}{
  language=Python,
  aboveskip=3mm,
  showstringspaces=false,
  columns=flexible,
  numbers=none,
  backgroundcolor=\color{backcolour},
  commentstyle=\color{codegreen},
 keywordstyle=\color{magenta},
    numberstyle=\tiny\color{codegray},
    stringstyle=\color{codepurple},
    basicstyle=\small\ttfamily,
    breakatwhitespace=false,         
    breaklines=false,                 
    captionpos=b,                    
    keepspaces=false,                 
    numbersep=5pt,                  
    showspaces=false,                
    showstringspaces=false,
    showtabs=false,                  
    tabsize=2,
    escapeinside=``
}
\definecolor{nima}{RGB}{1.0, 0.49, 0.0}
\definecolor{songcolor}{RGB}{191,191,191}
\definecolor{nimacolor}{RGB}{0.13, 0.67, 0.8}
\definecolor{aruncolor}{RGB}{51,255,51}

\newcommand{\add}[1]{\textcolor{blue}{\textbf{#1}}}

\newcommand{\moses}[1]{\textcolor{blue}{{\it [Moses: #1]}}}

\newmdenv[
  backgroundcolor=lightblue,
  linecolor=black,
  skipabove=10pt,
  skipbelow=10pt,
  linewidth=1pt,
  innertopmargin=6pt,
  innerbottommargin=6pt,
  innerleftmargin=6pt,
  innerrightmargin=6pt,
  frametitlebackgroundcolor=lightblue,
  frametitlefont=\bfseries,
]{findingbox} 

\begin{document}

\title{A Large-Scale Empirical Study of Quality Assurance Practices and Gaps in AI Agents}
\author{
\IEEEauthorblockN{Wuyang Dai}
\IEEEauthorblockA{
York University\\
ddai2002@my.yorku.ca
}
\and
\IEEEauthorblockN{Moses Openja}
\IEEEauthorblockA{
Polytechnique Montréal\\
openjamosesopm@gmail.com
}
\and
\IEEEauthorblockN{Jiho Shin}
\IEEEauthorblockA{
York University\\
shinjiho@yorku.ca
}
\and
\IEEEauthorblockN{Hung Viet Pham}
\IEEEauthorblockA{
York University\\
hvpham@yorku.ca
}
\and
\IEEEauthorblockN{Song Wang}
\IEEEauthorblockA{
York University\\
wangsong@yorku.ca
}
}

\maketitle

\begin{abstract}
Large language model (LLM)-based agents are increasingly used in software engineering, web automation, research assistance, personal productivity, and other domains. By integrating LLMs with planning, memory, retrieval, tool use, code execution, and interaction with the external environment, these agents can perform complex tasks with increasing autonomy. However, this autonomy also introduces new reliability, safety, and security risks, creating a need for systematic quality assurance (QA) practices tailored to agentic systems. 

In this paper, we present a large-scale empirical study of QA practices in 157 open-source LLM-based agent projects, each with at least 100 GitHub stars. We analyze repository documentation, source code, configuration files, tests, and structured review evidence to characterize how current projects address QA practices across four layers: execution surfaces, safeguards, testing artifacts, risk scenarios, and recurring QA gaps. 
Our results show that existing QA practices primarily focus on validating basic functionality and constraining high-risk actions. However, coverage remains fragmented across the four layers: the breadth of execution surfaces makes QA coverage difficult to trace, controls are not always applied consistently across equivalent routes, tests rarely target boundary, adversarial, or multi-step tool-use failures, and potential risk scenarios are seldom translated into end-to-end QA checks. 

Overall, our findings show that current agent QA practices provide only partial protection against high-risk agent actions and do not consistently ensure that model-influenced behavior is bounded across all execution surfaces. We argue that agent QA should move beyond feature-level checks toward end-to-end validation of whether agent workflows remain within intended boundaries when they combine untrusted inputs, model decisions, tools, persistent state, and external APIs.

\end{abstract}


\section{Introduction}
\label{sec:intro}

Large language model (LLM)-based agents are rapidly changing how software systems are designed, developed, and used~\cite{wang2023surveyagents,xi2023rise}. Unlike traditional LLM applications that primarily return direct responses, agents combine models with planning, memory, tool use, code execution, retrieval, browser interaction, file manipulation, and external services~\cite{yao2023react,karpas2022mrkl,shinn2023reflexion,wu2023autogen}. These capabilities support multi-step autonomy across software engineering~\cite{hong2023metagpt,qian2023chatdev}, browser and computer automation~\cite{debenedetti2024agentdojo}, research assistance~\cite{wang2023surveyagents,xi2023rise}, and security-oriented evaluation~\cite{ruan2023toolemu,zhang2025asb}.

This shift creates a quality-assurance challenge beyond response correctness. A conventional LLM application may produce an incorrect answer, a hallucinated explanation, or a misleading recommendation. 
An agent may instead execute shell commands, modify files, browse websites, call APIs, install packages, or delegate tasks to subagents. Model outputs can therefore influence external actions~\cite{ruan2023toolemu,debenedetti2024agentdojo,zhang2025asb}, with consequences such as corrupted files, privacy leakage, unsafe commands, insecure dependency installation, financial loss, compromised credentials, or harmful third-party interactions~\cite{greshake2023indirect,liu2023houyi,liu2024formalizing,owasp2025llm}. Agent QA must therefore ask whether model-influenced actions are bounded across interaction and execution surfaces.

Despite the rapid evolution of LLM-based agents and the growing body of work on agent architectures, benchmarks, and security, little is known about how quality assurance is actually engineered in the wild. It remains unclear which QA practices developers adopt, how safeguards and testing are integrated across different execution paths, and whether current repositories provide evidence of end-to-end assurance for model-mediated actions. Open-source repositories provide a unique opportunity to study these engineering practices because they expose testing infrastructure, safeguard implementations, configuration files, documentation, evaluation scripts, and deployment artifacts that collectively reflect observable QA activities during development. To fill this gap, we conduct a large-scale empirical study of 157 open-source LLM-based agent projects, systematically characterizing repository-visible QA practices across architecture, execution surfaces, safeguards, testing artifacts, and risk scenarios.
The corpus spans SWE automation (26.8\%), personal and multichannel assistants (19.7\%), coding assistants (19.1\%), research and information agents (13.4\%), browser/computer-use automation (12.1\%), and domain-specific agents (8.9\%).

{Because LLM-based agent behavior emerges from interactions among reasoning, planning, execution interfaces, external tools, and runtime safeguards, QA cannot be evaluated solely by inspecting the model outputs. Instead, it must consider how actions are introduced into architecture, exposed through execution surfaces, constrained by safeguards, validated through testing, and connected through potential risk scenarios.}
We therefore analyze repository-visible QA practices across four layers: \textit{execution surfaces}, \textit{safeguards}, \textit{testing artifacts}, and \textit{risk scenarios}. For each layer, we code what the repository makes inspectable, including action routes, controls, tests, and risk chains. These codes indicate visible QA practices in the repository; they do not show that a mechanism is effective after deployment. Section~\ref{sec:experiment} defines the coding rubric and review process.

Through this analysis, we identify recurring QA practices and uncover important QA gaps across the model-to-action workflow. Our results show that agent projects often expose broad execution surfaces and include controls such as approval, sandboxing, permission checks, redaction, and domain scoping. However, these controls are not always applied consistently across equivalent routes. Agent-specific tests for boundary, adversarial, or multi-step tool-use failures remain rare. The risk scenarios we report should therefore be read as candidate model-to-action situations that need boundary review, not as confirmed vulnerabilities. 
These findings suggest that QA for agents should extend beyond model-output correctness to systematic reasoning about the complete model-to-action workflow. In addition to asking whether the model gives the right answer, agent QA should ask whether model-influenced actions remain authorized, scoped, monitored, and tested when they pass through tools, plugins, browser state, permissions, runtime execution, and user-facing paths.

The remainder of this paper is organized as follows. Section~\ref{sec:bg} introduces the background and related work. Section~\ref{sec:experiment} presents the methodology and empirical study design. Section~\ref{sec:result} reports the results. Section~\ref{Threats} discusses threats to validity, and Section~\ref{sec:conclusion} concludes the paper.

\noindent \textbf{Data Availability:} \url{https://anonymous.4open.science/r/Quality-Assurance-Practices-and-Gaps-in-AI-Agents-85F2/} 

\section{Background and Related Work}
\label{sec:bg}

LLM-based agents combine model inference with memory, planning, tools, perception, and cooperation \cite{wang2023surveyagents,xi2023rise}. Prior work has introduced core patterns such as ReAct-style reasoning and acting \cite{yao2023react}, MRKL-style tool routing \cite{karpas2022mrkl}, reflection and search-based agents \cite{shinn2023reflexion,yao2023tree,zhou2024lats}, memory-centered agents \cite{park2023generative,wang2023voyager}, and multi-agent coordination frameworks \cite{wu2023autogen,li2023camel,hong2023metagpt,qian2023chatdev}. These architectures are usually discussed as capability mechanisms, but they also define trust boundaries: ReAct loops repeatedly mix observations with actions, tool routers map model decisions to external functions, memory systems persist prior context, and multi-agent systems delegate work across roles or workers. We use this literature to code repository architecture, but our focus is narrower: how these architectures expose and constrain safety- and security-relevant execution paths in real projects.

Prior security work motivates this focus. Prompt-injection studies show that untrusted content can redirect model behavior, especially when webpages, documents, emails, issues, or repository files enter agent context \cite{greshake2023indirect,liu2023houyi,liu2024formalizing}. This risk becomes more concrete in agents because injected instructions can be followed by tool calls, file access, browser actions, or API requests. Agent benchmarks such as ToolEmu, AgentDojo, and Agent Security Bench evaluate attacks and defenses in tool-using environments \cite{ruan2023toolemu,debenedetti2024agentdojo,zhang2025asb}, while OWASP's LLM guidance highlights prompt injection, excessive agency, insecure output handling, sensitive information disclosure, and supply-chain risk \cite{owasp2025llm}. Recent MCP and secure-agent studies further examine MCP security, capability-oriented defenses, and cross-tool state leakage or pollution \cite{hasan2025mcpfirstglance,yang2025mcpsecbench,debenedetti2025camel,li2025dissonances}. Together, these works identify important risk classes, but they do not directly measure how often open-source agent repositories expose the corresponding tools, policies, tests, and configuration paths.

Empirical OSS security research mines repositories, packages, and static-analysis warnings to understand security practice in the wild \cite{aloraini2019sast,zahan2022weaklinks}; recent agent-testing work studies developer-written test functions across agent frameworks and applications \cite{hasan2025testingpractices}. {Our study complements
this body of work by providing a repository-level perspective on how QA is engineered in practice. 
Whereas prior work primarily evaluates agent capabilities, benchmark performance, or individual attacks, our study systematically analyzes repository-visible engineering artifacts that provide evidence of QA in open-source LLM-based agents.} We take a broader view in which tests are one evidence source alongside code, configuration, documentation, safeguards, execution surfaces, and structured risk observations. This distinction matters because a repository may document a safeguard without testing it, test a permission component without applying it to every interface, or expose a risky surface without containing a confirmed vulnerability. 

\section{Empirical Study Design}
\label{sec:experiment}

\begin{figure}[t]
  \centering
  \includegraphics[width=\columnwidth]{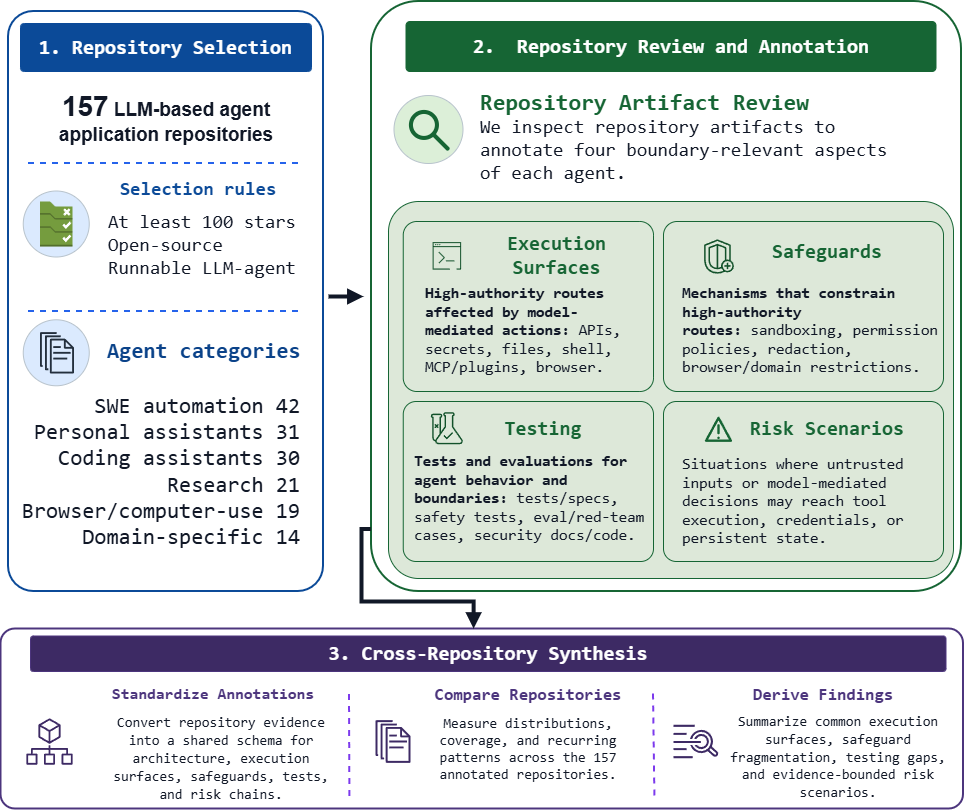}
  \caption{Workflow of our empirical study.}
  \label{fig:analysis-workflow}
\end{figure}

Figure~\ref{fig:analysis-workflow} illustrates the overall workflow of our study, which consists of three stages: repository selection, repository review and annotation, and cross-repository synthesis. {During the review stage, we apply a four-layer coding rubric, developed from the literature reviewed in Section~\ref{sec:bg} and our initial exploration of selected repositories, to annotate repository artifacts as execution-surface, safeguard, testing-artifact, and risk-scenario records. These layers separate where agents can act, how actions are controlled, how behavior is tested, and which situations require boundary review.} {During the synthesis stage, we aggregate these coded records to derive corpus-level statistics, QA practice families, and undercovered QA problems.}

\subsection{Terminology}
\label{sec:terminology}

{To avoid ambiguity, we use three terms consistently. \textit{Repository artifacts} are the inspected sources, including documentation, source code, configuration files, manifests, tests, examples, and deployment artifacts. \textit{Coded records} are the structured annotations derived from those artifacts, such as execution-surface records, safeguard records, testing-artifact records, and risk-scenario records. \textit{Analysis layers} are the four categories used to organize those records: execution surfaces, safeguards, testing artifacts, and risk scenarios.}

\subsection{Repository Selection}
\label{3.1}

For our analysis, we target popular LLM-based agent applications whose repositories provide enough artifacts to examine QA-relevant behavior and contain executable entry points or package manifests.
We focus on public GitHub projects with at least 100 stars as of May 2026, so that the corpus reflects actively visible projects with enough adoption and repository artifacts for systematic review.
We first collected candidate repositories using agent-focused search terms, including ``AI agent'', ``LLM agent'', ``autonomous agent'', ``coding agent'', ``browser agent'', ``computer use agent'', and ``MCP agent'', supplemented by curated project lists and snowballing from related repositories.
A rerun of these queries on 2026-05-21 returned 5,852 summed raw hits before overlap removal and eligibility filtering.
We then excluded repositories that did not satisfy three inclusion criteria: the project must integrate an LLM or model-compatible API, contain an executable entry point or package manifest, and expose at least one model-mediated action surface, such as tool calling, shell/process execution, file editing, browser automation, MCP/plugin integration, workflow execution, or persistent agent state.
After applying these criteria, our final corpus contains 157 open-source LLM-based agent projects cloned from GitHub.
The filtered manifest records each repository's name, URL, local directory, analyzed commit, timestamps, project-purpose category, interface form, and star count. We treat the corpus as a fixed sample of popular agent applications, not an exhaustive census of all GitHub agent projects.
Table~\ref{tab:corpus-category} reports the distribution of agent categories.

\begin{table}[t!]
  \centering
  \small
  \caption{Corpus composition by project category.}
  \label{tab:corpus-category}
  \begin{tabularx}{0.85\linewidth}{Xrr}
    \toprule
    Category & Repos. & Share \\
    \midrule
    SWE automation agents & 42 & 26.8\% \\
    Personal / multichannel assistants & 31 & 19.7\% \\
    Coding assistants & 30 & 19.1\% \\
    Research / information agents & 21 & 13.4\% \\
    Browser / computer-use automation & 19 & 12.1\% \\
    Domain-specific agents & 14 & 8.9\% \\ \hline
    \textbf{Total} & 157 & -\\
    \bottomrule
  \end{tabularx}
\end{table}

\subsection{Repository Review and Annotation} 
\label{sec:3.2}
To systematically characterize repository-visible QA evidence, we first present the coding rubric and then describe the annotation procedure.
\subsubsection{Coding Rubric}
\label{3.2.1}



{The coding rubric consists of four analysis layers: \textit{execution surfaces}, \textit{safeguards}, \textit{testing artifacts}, and \textit{risk scenarios}. We use the rubric to turn repository artifacts into coded records that answer four operational questions. Execution-surface records identify the interfaces and action channels through which an agent can receive requests or affect external systems, following work on tool use and side-effecting agent environments~\cite{schick2023toolformer,ruan2023toolemu,debenedetti2024agentdojo,zhang2025asb}. Safeguard records identify controls that authorize, scope, monitor, or block agent actions, following prompt-injection and agent-security work on tool, browser, and policy boundaries~\cite{greshake2023indirect,liu2024formalizing,zhan2024injecagent,debenedetti2024agentdojo,owasp2025llm}. Testing-artifact records identify tests, evaluations, benchmarks, and failure-oriented checks for agent behavior under tasks, tools, and environments~\cite{liu2024agentbench,ruan2023toolemu,debenedetti2024agentdojo,zhang2025asb}. Risk-scenario records identify situations where agent behavior may reach sensitive actions, assets, or persistent state and therefore require boundary review~\cite{greshake2023indirect,liu2024formalizing,zhan2024injecagent,debenedetti2024agentdojo,zhang2025asb,yang2025mcpsecbench,li2025dissonances,owasp2025llm}; they are not confirmed vulnerabilities.}

\begin{table}[t]
\centering
\caption{QA-relevant analysis layers used in the study.}
\label{tab:qa-dimensions}
\resizebox{\columnwidth}{!}{%
\begin{tabular}{p{0.18\linewidth} p{0.56\linewidth} p{0.19\linewidth}}
\toprule
\textbf{Layer} & \textbf{Description} & \textbf{References} \\
\midrule
\textit{Execution Surfaces}
& Where an agent can be invoked or can act, including UI, CLI, API, browser, filesystem, shell, plugin, and subagent paths.
& \cite{schick2023toolformer,yang2024swe,ruan2023toolemu,debenedetti2024agentdojo,hasan2025mcpfirstglance} \\

\textit{Safeguards}
& Mechanisms that constrain, validate, or monitor model-mediated actions, such as approval, policy, sandboxing, redaction, and domain scoping.
& \cite{greshake2023indirect,liu2024formalizing,zhan2024injecagent,debenedetti2024agentdojo,debenedetti2025camel,owasp2025llm} \\

\textit{Testing Artifacts}
& Tests, evaluations, benchmarks, reproduction scripts, security tests, and failure-oriented checks.
& \cite{hasan2025testingpractices,liu2024agentbench,ruan2023toolemu,debenedetti2024agentdojo,zhang2025asb} \\

\textit{Risk Scenarios}
& Situations where untrusted context or agent behavior may reach sensitive actions, assets, or persistent state.
& \cite{greshake2023indirect,liu2024formalizing,zhan2024injecagent,debenedetti2024agentdojo,zhang2025asb,yang2025mcpsecbench,li2025dissonances,owasp2025llm} \\
\bottomrule
\end{tabular}
}
\end{table}


\subsubsection{Annotation Process}
\label{3.2.2}

Before applying the coding rubric, reviewers first familiarized themselves with each repository by reading README files, project documentation, configuration files, and directory structure to understand the project purpose, agent workflow, supported tools, deployment model, and permission assumptions. Reviewers then inspected repository artifacts related to agent execution, including documentation, source code, configuration files, manifests, tests, examples, deployment artifacts, and CI workflows.These artifacts were then coded using the four-layer rubric, recording project purpose, execution surfaces, safeguards, testing artifacts, and risk-scenario candidates.

{Our annotation followed three guidelines informed by qualitative-analysis practices in empirical software engineering~\cite{seaman1999qualitative,cruzes2011thematic}. First, every label had to point to a repository-visible artifact. Second, capability and QA labels were allowed to co-exist but captured different properties of the artifact: shell, browser, MCP, or plugin access was coded as a capability, while approval prompts, permission checks, sandboxing, redaction, monitoring, or tests were coded as QA mechanisms associated with that capability.} 
{Third, QA labels record the presence of reviewable mechanisms or testing intent, not validated runtime effectiveness; for example, a sandbox setting or test path shows that the repository contains a relevant QA mechanism, but we do not claim that the mechanism prevents all corresponding failures in deployed use.} 

We developed the label set through open coding and annotation cards. Reviewers first inspected a diverse subset of repositories and wrote cards for recurring behaviors, recording the repository, artifact/path, behavior summary, tentative layer, and uncertainty note. Related cards were grouped and refined into the final labels; for example, a plugin artifact could be an execution-surface record when it added a model-callable extension channel, a safeguard record when it implemented permission checks, or a risk-scenario record when it linked plugin authority to untrusted context.


\subsection{Research Questions}
\label{sec:rq}

We organize the study around two research questions. RQ1 examines coded records across four analysis layers, and RQ2 identifies the QA problems that remain undercovered within and across those layers.

\begin{itemize}
\item \textbf{RQ1: What visible QA practice appears in open-source LLM-based agent projects?}
To answer this question, we examine four QA layers:
\begin{itemize}
 \item \textbf{RQ1.1: QA Challenges and Practices at the Execution surfaces.} How do projects expose, separate, document, or qualify the entry points and action channels through which agents can affect external systems?

 \item \textbf{RQ1.2: QA Challenges and Practices at the Safeguards.}What safeguards are visible for constraining, validating, or monitoring model-mediated behavior before it reaches sensitive actions?

 \item \textbf{RQ1.3: QA Challenges and Practices at the Testing Layer.} What testing and evaluation artifacts are visible through tests, benchmarks, reproduction paths, robustness checks, security tests, or failure-oriented paths?

\item \textbf{RQ1.4: QA Challenges and Practices in Risk scenarios.} What recurring situations allow agent behavior to reach sensitive actions, assets, or persistent state?

 \end{itemize}

\item \textbf{RQ2: What QA problems remain undercovered in current open-source LLM-based agent projects?}

This question identifies missing, fragmented, or weak QA coverage across the four layers, including gaps in end-to-end workflow validation, cross-interface safeguard consistency, robustness and security testing, environment-aware validation, and documentation of bounded actions and trusted assumptions.
\end{itemize}

\section{Results}
\label{sec:result}

This section reports an evidence-led synthesis for the two research questions. 




\subsection{RQ1.1: QA Challenges and Practices at the Execution Surface 
Layer}


The execution surface is where an agent can affect external systems. The QA for this layer is therefore not only whether a surface exists, but whether the project can keep those surfaces safe. We derived the QA challenges and QA-practice families through manual review of repository-visible model-to-action paths, configuration paths, tool interfaces, and controls attached to those paths. Table~\ref{tab:rq12-issue-practice} summarizes the main types of QA challenges and QA practices projects use to address them; percentages in the third column use the QA-challenge project count as the denominator.

\begin{table*}[t]
  \centering
  \scriptsize
  \caption{Execution-surface QA challenges and QA practices.}
  \label{tab:rq12-issue-practice}
  \setlength{\tabcolsep}{3pt}
  \renewcommand{\arraystretch}{1.08}
  \begin{tabularx}{\textwidth}{>{\raggedright\arraybackslash}p{0.30\textwidth}r>{\raggedright\arraybackslash}X}
    \toprule
    QA challenge & Projects & QA practice\\
    \midrule
    Inconsistent control across overlapping action paths & 152 (96.8\%) & Approval and authorization controls: 142 (93.4\%); sandboxing and isolation mechanisms: 106 (69.7\%); permission and policy controls: 101 (66.4\%) \\
    Outbound network/API effects beyond the project boundary & 156 (99.4\%) & Approval and authorization controls: 143 (91.7\%); permission and policy controls: 102 (65.4\%) \\
    Credential leakage through configuration & 155 (98.7\%) & Permission and policy controls: 102 (65.8\%); secret management and redaction: 99 (63.9\%) \\
    Workspace escape through file access & 153 (97.5\%) & Sandboxing and isolation mechanisms: 107 (69.9\%); permission and policy controls: 101 (66.0\%); secret management and redaction: 98 (64.1\%) \\
    Host process and environment modification through commands & 148 (94.3\%) & Approval and authorization controls: 137 (92.6\%); sandboxing and isolation mechanisms: 103 (69.6\%); permission and policy controls: 98 (66.2\%) \\
    Extension-driven authority expansion & 127 (80.9\%) & Approval and authorization controls: 119 (93.7\%); MCP/plugin registration and configuration: 104 (81.9\%); permission and policy controls: 97 (76.4\%); secret management and redaction: 88 (69.3\%) \\
    Unreviewed runtime code execution & 123 (78.3\%) & MCP/plugin registration and configuration: 93 (75.6\%); sandboxing and isolation mechanisms: 89 (72.4\%); permission and policy controls: 87 (70.7\%) \\
    Browser session and cross-domain exposure & 119 (75.8\%) & Approval and authorization controls: 110 (92.4\%); browser and domain-scoping controls: 87 (73.1\%); permission and policy controls: 86 (72.3\%); secret management and redaction: 80 (67.2\%) \\
    \bottomrule
  \end{tabularx}
\end{table*}

\subsubsection{QA Challenges}

\begin{itemize}[leftmargin=*]
  \item \textbf{Inconsistent control across overlapping action paths} occurs when a project exposes several routes to similar authority, so a control on one route may not cover another route. For example, \texttt{google-gemini/gemini-cli} exposes CLI, shell, filesystem, browser, web, and MCP paths, which makes cross-path consistency a central QA challenge.
  \item \textbf{Outbound network/API effects beyond the project boundary} occur when model-mediated behavior can send data or trigger remote actions outside the repository. For example, \texttt{CherryHQ/cherry-studio} exposes outbound provider, MCP, search, browser, and local API channels.
  \item \textbf{Credential leakage through configuration} occurs when provider keys, OAuth stores, browser profiles, or MCP environment variables become reachable by tools or subprocesses. For example, \texttt{google-gemini/gemini-cli} supports configured MCP environments and provider settings that have to be kept away from logs and tool outputs.
  \item \textbf{Workspace escape through file access} occurs when file tools, downloads, skills, or hooks can read or write outside the intended workspace. For example, \texttt{OpenHands/OpenHands} has workspace, sandbox, plugin, hook, and persisted-setting paths around file access.
  \item \textbf{Host process and environment modification through commands} occurs when model-selected commands can change host process state, environment variables, or workspace contents. For example, \texttt{openai/codex} exposes shell execution paths whose authority depends on approval and sandbox mode.
  \item \textbf{Extension-driven authority expansion} occurs when MCP servers, plugins, or skills add new model-callable tools and local subprocesses. For example, \texttt{aaif-goose/goose} treats MCP extensions as a primary capability.
  \item \textbf{Unreviewed runtime code execution} occurs when configured tools, imported modules, JavaScript, Python snippets, or runtime-loaded plugins execute logic outside the fixed application path. For example, \texttt{google-gemini/gemini-cli} supports extension and MCP configuration that can alter the set of callable runtime capabilities.
  \item \textbf{Browser session and cross-domain exposure} occurs when browser automation can interact with signed-in sessions, page scripts, cookies, uploads, downloads, or domains. For example, \texttt{browser-use/browser-use} exposes browser navigation and page interaction over persistent browser state.
\end{itemize}

\subsubsection{QA Practice}
\begin{itemize}[leftmargin=*]
  \item \textbf{Approval and authorization controls} require an explicit grant before a route performs a high-authority action or exposes sensitive API state. For example, \texttt{CherryHQ/cherry-studio} protects sensitive local API routes with bearer authentication, while \texttt{google-gemini/gemini-cli} uses confirmation and trust modes around interactive actions. This practice addresses inconsistent control across overlapping action paths, outbound network/API effects, host process modification through commands, extension-driven authority expansion, and browser session exposure.
  \item \textbf{Sandboxing and isolation mechanisms} run actions inside a bounded workspace, container, or restricted execution mode so file writes, command effects, and runtime-loaded code are separated from the host. For example, \texttt{OpenHands/OpenHands} separates workspace state and sandbox execution, and \texttt{openai/codex} routes shell commands through sandbox policy. This practice addresses overlapping action paths, workspace escape through file access, host process modification, and unreviewed runtime code execution.
  \item \textbf{Permission and policy controls} encode rules about which actions, paths, tools, routes, or modes are allowed before execution. For example, \texttt{1Panel-dev/MaxKB} uses token permissions and tool filtering around custom Python tools and generated MCP servers. This practice addresses network/API effects, credential leakage, filesystem boundaries, commands, extensions, dynamic code, and browser actions by putting the allowed action set into a reviewable policy layer.
  \item \textbf{Secret management and redaction} limit where credentials can be stored, inherited, logged, or returned through a model/tool path. For example, \texttt{google-gemini/gemini-cli} includes logging redaction and environment sanitization around tool and provider flows. This practice directly addresses credential leakage through configuration and also supports filesystem, plugin, dynamic-code, and browser-session challenges where secrets may be inherited or exposed.
  \item \textbf{MCP/plugin registration and configuration} makes extension points explicit by listing configured servers, tools, commands, or generated plugin interfaces. For example, \texttt{aaif-goose/goose} pairs MCP extension support with permission, confirmation, and server-auth mechanisms. This practice addresses extension-driven authority expansion and unreviewed runtime code execution, and it also helps review overlapping action paths when plugin tools duplicate shell, file, browser, or network authority.
  \item \textbf{Browser and domain-scoping controls} restrict which domains, pages, scripts, uploads, downloads, or session data a browser agent can use. For example, \texttt{browser-use/browser-use} includes allowed/prohibited domain settings, sensitive-data filtering, IP blocking, storage-state handling, and upload-path gating. This practice addresses browser session and cross-domain exposure, and it also reduces credential leakage and outbound network/API effects when signed-in pages or uploaded files contain sensitive data.
\end{itemize}

\mybox{\textbf{Answer to RQ1.1.} QA at the \textit{execution-surface} layer primarily concerns authority boundaries: model decisions can reach credentials, files, commands, plugins, browser state, and external APIs through multiple routes, yet route-level controls are often unevenly applied.}

\subsection{{RQ1.2: QA Challenges and Practices at the Safeguard Layer}}

The safeguard layer examines whether projects put clear limits around high-authority agent actions. The main potential issue is incomplete control coverage: a project may protect one route to shell, file, browser, plugin, or secret access while another route to similar authority is less clearly bounded. Table~\ref{tab:rq13-issue-practice} summarizes the main safeguard issues and the safeguard practice families projects use to address them.

\begin{table*}[t]
  \centering
  \scriptsize
  \caption{{Safeguard QA challenges and QA practice families.}}
  \label{tab:rq13-issue-practice}
  \setlength{\tabcolsep}{3pt}
  \renewcommand{\arraystretch}{1.08}
  \begin{tabularx}{\textwidth}{>{\raggedright\arraybackslash}p{0.30\textwidth}r>{\raggedright\arraybackslash}X}
    \toprule
    QA challenge & Projects & QA practice \\
    \midrule
    Privileged command execution needs approval and containment & 148 (94.3\%) & Approval and authorization controls: 137 (92.6\%); sandboxing and isolation mechanisms: 103 (69.6\%); permission and policy controls: 98 (66.2\%); secret management and redaction: 94 (63.5\%) \\
    File and workspace access needs boundary control & 153 (97.5\%) & Sandboxing and isolation mechanisms: 107 (69.9\%); permission and policy controls: 101 (66.0\%); secret management and redaction: 98 (64.1\%) \\
    Secrets and credentials need scoped handling & 155 (98.7\%) & Approval and authorization controls: 142 (91.6\%); sandboxing and isolation mechanisms: 107 (69.0\%); permission and policy controls: 102 (65.8\%); secret management and redaction: 99 (63.9\%) \\
    Plugin and MCP authority needs extension control & 144 (91.7\%) & Approval and authorization controls: 135 (93.8\%); MCP/plugin registration and configuration: 105 (72.9\%); permission and policy controls: 98 (68.1\%); secret management and redaction: 93 (64.6\%) \\
    Browser session access needs domain and data scoping & 119 (75.8\%) & Approval and authorization controls: 110 (92.4\%); browser and domain-scoping controls: 87 (73.1\%); permission and policy controls: 86 (72.3\%); secret management and redaction: 80 (67.2\%) \\
    \bottomrule
  \end{tabularx}
\end{table*}

\subsubsection{QA Challenges}

\begin{itemize}[leftmargin=*]
  \item \textbf{Privileged command execution needs approval and containment} because model-selected commands can change host process state, environment variables, or workspace contents. For example, \texttt{openai/codex} exposes shell execution paths whose authority depends on approval and sandbox mode.
  \item \textbf{File and workspace access needs boundary control} because file tools, downloads, skills, or hooks can read or write outside the intended workspace. For example, \texttt{OpenHands/OpenHands} separates workspace state and sandbox execution, which makes the workspace boundary a central review point.
  \item \textbf{Secrets and credentials need scoped handling} because provider keys, OAuth tokens, browser profiles, and MCP environment variables can become reachable by tools, subprocesses, logs, or provider calls. For example, \texttt{google-gemini/gemini-cli} combines provider configuration and MCP environments with logging and redaction paths.
  \item \textbf{Plugin and MCP authority needs extension control} because plugins and MCP servers can add model-callable tools, local subprocesses, and inherited environment access. For example, \texttt{aaif-goose/goose} treats MCP/plugin support as a primary capability that expands the set of callable actions.
  \item \textbf{Browser session access needs domain and data scoping} because browser agents can interact with signed-in sessions, page scripts, downloads, uploads, cookies, or domains. For example, \texttt{browser-use/browser-use} exposes browser navigation and page interaction over browser state that may persist across tasks or domains.
\end{itemize}

\subsubsection{QA Practices}

\begin{itemize}[leftmargin=*]
  \item \textbf{Approval and authorization controls} require an explicit grant before privileged execution or access to sensitive state. For example, \texttt{openai/codex} routes shell commands through approval policy, and \texttt{aaif-goose/goose} uses confirmation and server-auth mechanisms around extension paths. This practice addresses privileged command execution, plugin/MCP authority, browser session access, and secrets or credentials that should not be exposed without authorization.
  \item \textbf{Sandboxing and isolation mechanisms} run actions inside a bounded workspace, container, or restricted execution mode so command effects, file writes, and runtime-loaded code are separated from the host. For example, \texttt{OpenHands/OpenHands} separates workspace state from sandbox execution, while \texttt{openai/codex} supports sandbox policy around shell execution. This practice addresses privileged command execution, file/workspace boundary control, secrets and credentials, and plugin or runtime paths that may inherit host authority.
  \item \textbf{Permission and policy controls} encode rules about which commands, files, tools, routes, or modes are allowed before execution. For example, \texttt{1Panel-dev/MaxKB} uses token permissions and tool filtering around custom Python tools and generated MCP servers. This practice addresses command execution, file/workspace access, credential handling, plugin/MCP authority, and browser session access by making the allowed action set reviewable.
  \item \textbf{Secret management and redaction} limit where credentials can be stored, inherited, logged, or returned through a model/tool path. For example, \texttt{google-gemini/gemini-cli} includes logging redaction and environment sanitization around tool and provider flows. This practice directly addresses secrets and credentials, and it also supports file/workspace, plugin/MCP, command, and browser-session challenges where sensitive values can be inherited or exposed.
  \item \textbf{MCP/plugin registration and configuration} makes extension authority explicit by listing configured servers, tools, commands, or generated plugin interfaces. For example, \texttt{aaif-goose/goose} pairs MCP extension support with permission, confirmation, and server-auth mechanisms. This practice addresses plugin/MCP authority and helps review command, file, credential, and browser challenges introduced by extension-provided tools.
  \item \textbf{Browser and domain-scoping controls} restrict which domains, pages, scripts, uploads, downloads, or session data a browser agent can use. For example, \texttt{browser-use/browser-use} includes allowed/prohibited domain settings, sensitive-data filtering, IP blocking, storage-state handling, and upload-path gating. This practice addresses browser session access and also reduces credential exposure and file/upload risks when signed-in pages or local files contain sensitive data.
\end{itemize}

\mybox{\textbf{Answer to RQ1.2.} For \textit{safeguard} layer, QA practices are widely present, but the key challenge is context-sensitive coverage: the same agent capability may be acceptable in one setting and risky in another, so safeguards need to account for how, where, and under what authority the action is reached.} 

\subsection{RQ1.3: QA Challenges and Practices at the Testing Layer}

Software testing itself is a conventional QA mechanism. 
{From the repository review, we grouped testing artifacts into four testing-practice categories.}
Across the 157 projects, the counts are: 137 (87.3\%) contain conventional tests or specs, 95 (60.5\%) contain security/safety-oriented tests, 78 (49.7\%) contain evaluation, benchmark, adversarial, or red-team-like paths, and 8 (5.1\%) contain prompt-injection, adversarial, jailbreak, or red-team testing paths.

\textbf{Conventional tests and specs.} These tests address ordinary implementation regressions in tool wrappers, API routes, storage, settings, UI helpers, provider clients, and parsers. They are useful because many agent failures still begin as ordinary software bugs: a tool argument is parsed incorrectly, a session setting is not persisted, or a file operation returns the wrong result. For example, \texttt{OpenHands/OpenHands} tests sandbox cleanup behavior, including no-sandbox cases, pagination, stale sandbox cleanup, pause failures, invalid cleanup limits, and sorting by creation time; it also tests secret-store initialization, serialization, provider-token handling, and frontend role-permission checks. \texttt{browser-use/browser-use} tests CLI sessions, filesystem integration, action-loop detection, coordinate clicking, browser navigation, and setup-command handling.

\textbf{Security/safety-oriented tests.} These tests address boundary regressions around high-authority actions, especially shell execution, filesystem access, browser sessions, domain policy, secrets, approval, sandboxing, and tool permissions. They are more directly agent-specific because the failure condition is not just that a function returns the wrong value, but that an agent action crosses a boundary it should not cross. For example, \texttt{browser-use/browser-use} tests that prohibited domains are blocked, allowed-domain lists override broad access, authentication text in a URL does not bypass domain checks, private or localhost IP addresses can be blocked, password fields are removed from DOM snapshots, and sensitive values are filtered from action results. \texttt{google-gemini/gemini-cli} tests shell-safety cases such as preferring file-write tools over shell commands for file creation, avoiding destructive shell commands, recovering when sandbox permissions are insufficient, and masking long tool outputs before they are reused.

\textbf{Evaluation and red-team-like practices.} These practices address behavior that only appears after the model chooses actions over one or more steps. They include task evals, benchmark runners, replayable response fixtures, adversarial prompts, and red-team scenarios. For example, \texttt{google-gemini/gemini-cli} uses evals for automated tool use, plan mode, subtask delegation, browser-agent workflows, background processes, memory behavior, and eval regression; these checks ask whether the agent chooses the expected tool sequence rather than only whether a helper function works. \texttt{Alibaba-NLP/DeepResearch} uses JSONL eval data and evaluation scripts for deep-search and web-agent workflows. \texttt{cordum-io/cordum} includes red-team scenarios that exercise approval, policy, tenant isolation, validation, and limit handling.

\textbf{Prompt-injection and adversarial tests.} These tests address the potential issue that untrusted pages, files, prompts, skills, or MCP responses can steer the agent away from the user's intent or policy. They are a specialized form of security/safety or red-team testing. For example, \texttt{Yeachan-Heo/oh-my-claudecode} tests that subagent prompts prepend a fixed system header, preserve the order of system, file, and user context, reject context file paths containing newlines, carriage returns, null bytes, or path traversal, and reject invalid agent-role names. \texttt{NousResearch/hermes-agent} includes prompt-injection and red-teaming tests around scheduled tools and bundled skills.

\mybox{\textbf{Answer to RQ1.3.} Testing remains dominated by conventional regression tests; agent-specific tests for boundary violations, untrusted-context attacks, and multi-step tool behavior are much less systematic.}

\subsection{RQ1.4: QA Challenges and Practices in Risk Scenarios Layer}

The risk-scenario layer is organized around risk-chain templates. Here, the unit of analysis is the end-to-end path rather than the individual surface or control: behaviors that are ordinary in isolation, such as reading files, invoking tools, reusing state, loading extensions, or using browser sessions, may become risky when connected through model decisions or untrusted context before reaching a sensitive action or asset. Table~\ref{tab:rq15-issue-practice} reports risk-chain QA challenges and the QA practices used in projects showing each chain.

\begin{table*}[t]
  \centering
  \scriptsize
  \caption{Risk-chain QA challenges and QA practices.}
  \label{tab:rq15-issue-practice}
  \setlength{\tabcolsep}{3pt}
  \renewcommand{\arraystretch}{1.08}
  \begin{tabularx}{\textwidth}{>{\raggedright\arraybackslash}p{0.30\textwidth}r>{\raggedright\arraybackslash}X}
    \toprule
    Risk-chain QA challenge & Projects & QA practices within risk-chain projects \\
    \midrule
    Secret-to-tool chain & 149 (94.9\%) & Approval and authorization controls: 137 (91.9\%); sandboxing and isolation mechanisms: 104 (69.8\%); permission and policy controls: 100 (67.1\%); secret management and redaction: 96 (64.4\%) \\
    Workspace-to-persistent-state chain & 128 (81.5\%) & Sandboxing and isolation mechanisms: 90 (70.3\%); permission and policy controls: 84 (65.6\%); secret management and redaction: 82 (64.1\%) \\
    Extension-to-authority chain & 137 (87.3\%) & Approval and authorization controls: 129 (94.2\%); MCP/plugin registration and configuration: 101 (73.7\%); permission and policy controls: 94 (68.6\%); secret management and redaction: 91 (66.4\%) \\
    Mode-switch-to-privilege chain & 123 (78.3\%) & Approval and authorization controls: 114 (92.7\%); sandboxing and isolation mechanisms: 84 (68.3\%); permission and policy controls: 82 (66.7\%) \\
    Runtime-code chain & 99 (63.1\%) & Approval and authorization controls: 92 (92.9\%); MCP/plugin registration and configuration: 75 (75.8\%); sandboxing and isolation mechanisms: 73 (73.7\%); permission and policy controls: 69 (69.7\%) \\
    Model-to-process chain & 131 (83.4\%) & Approval and authorization controls: 123 (93.9\%); sandboxing and isolation mechanisms: 91 (69.5\%); permission and policy controls: 91 (69.5\%); secret management and redaction: 84 (64.1\%) \\
    Session-to-action chain & 60 (38.2\%) & Approval and authorization controls: 53 (88.3\%); browser and domain-scoping controls: 50 (83.3\%); permission and policy controls: 44 (73.3\%); secret management and redaction: 38 (63.3\%) \\
    Untrusted-context-to-action chain & 45 (28.7\%) & Approval and authorization controls: 39 (86.7\%); permission and policy controls: 27 (60.0\%); secret management and redaction: 25 (55.6\%); browser and domain-scoping controls: 24 (53.3\%) \\
    \bottomrule
  \end{tabularx}
\end{table*}

\subsubsection{QA Challenges}

\begin{itemize}[leftmargin=*]
  \item \textbf{Secret-to-tool chain.} A stored key, OAuth token, browser profile, or MCP environment variable becomes part of a model-mediated tool path and may then flow into logs, subprocesses, provider calls, or tool outputs. For example, \texttt{google-gemini/gemini-cli} combines MCP environments, tool arguments, errors, and provider-facing content, so the review question is whether secret scope is preserved across the whole path.
  \item \textbf{Workspace-to-persistent-state chain.} A model-selected file operation, hook, plugin, or subagent crosses from the task workspace into persisted settings, stored state, or secret storage. For example, \texttt{OpenHands/OpenHands} connects workspace state, settings, hooks, plugins, and secret storage, so the review question is whether the intended workspace boundary still holds after state is reused.
  \item \textbf{Extension-to-authority chain.} A configured plugin, MCP server, or skill adds a new model-callable action that inherits local subprocess, network, file, or environment authority. For example, \texttt{aaif-goose/goose} treats MCP extensions as first-class capabilities, so the review question is whether newly added tools inherit the same trust and permission assumptions as built-in tools.
  \item \textbf{Mode-switch-to-privilege chain.} An action that is checked in an interactive path is reachable through another mode, such as an automatic mode, setup hook, API route, headless path, or alternate tool. For example, \texttt{openai/codex} includes multiple shell-related modes and control paths, so the review question is whether the same privileged action remains bounded after the execution mode changes.
  \item \textbf{Runtime-code chain.} Configuration, imported modules, JavaScript, Python snippets, or runtime-loaded plugins introduce executable logic after the fixed application code has been reviewed. For example, \texttt{google-gemini/gemini-cli} combines extension and MCP configuration with runtime capability changes, so the review question is whether loaded code is subject to the same review boundary as built-in code.
  \item \textbf{Model-to-process chain.} A model-selected action reaches a host process, environment variable, background job, or workspace mutation. For example, \texttt{openai/codex} exposes command execution through more than one control path, so the review question is whether command authority is bounded at the action level rather than only at one interface.
  \item \textbf{Session-to-action chain.} Browser state, cookies, storage, uploads, downloads, or signed-in pages persist across domains or tasks and later influence an agent action. For example, \texttt{browser-use/browser-use} connects navigation, page interaction, storage state, and uploads, so the review question is whether session state remains scoped to the intended task and domain.
  \item \textbf{Untrusted-context-to-action chain.} A page, file, prompt, tool output, or MCP response influences a later tool call, command, file operation, or data disclosure. For example, \texttt{Yeachan-Heo/oh-my-claudecode} handles subagent context files, so the review question is whether untrusted context can move from input material into later privileged actions.
\end{itemize}

\subsubsection{QA Practices}
\leavevmode\par\smallskip

QA practices in this layer can be grouped into two broad categories. 

\begin{itemize}

   \item \noindent\textbf{Chain-interruption practices} stop a risky path at a critical transition point, such as when a model-mediated workflow reaches a privileged action, host execution, or sensitive data.
\begin{itemize}[leftmargin=*]
  \item \textbf{Authorize the transition to a privileged action.} Approval and authorization controls interrupt a chain when a model-mediated path reaches a sensitive action. This practice appears in 137 of 149 secret-to-tool chains, 129 of 137 extension-to-authority chains, 114 of 123 mode-switch-to-privilege chains, 123 of 131 model-to-process chains, 53 of 60 session-to-action chains, and 39 of 45 untrusted-context-to-action chains. For example, \texttt{openai/codex} uses an approval policy around command execution paths, and \texttt{aaif-goose/goose} uses confirmation and server-auth mechanisms around extension paths.
  \item \textbf{Contain execution or state effects.} Sandboxing and isolation mechanisms interrupt a chain by preventing a tool, command, runtime module, or workspace operation from inheriting unrestricted host authority. This practice appears in 104 of 149 secret-to-tool chains, 90 of 128 workspace-to-persistent-state chains, 84 of 123 mode-switch-to-privilege chains, 73 of 99 runtime-code chains, and 91 of 131 model-to-process chains. For example, \texttt{OpenHands/OpenHands} separates workspace state from sandbox execution, while \texttt{openai/codex} applies sandbox policy to shell execution.
  \item \textbf{Stop secret propagation.} Secret management and redaction interrupt a chain by preventing credentials from being inherited, logged, returned through tool output, or sent into provider-facing content. This practice appears in 96 of 149 secret-to-tool chains, 82 of 128 workspace-to-persistent-state chains, 91 of 137 extension-to-authority chains, 84 of 131 model-to-process chains, 38 of 60 session-to-action chains, and 25 of 45 untrusted-context-to-action chains. For example, \texttt{google-gemini/gemini-cli} includes logging redaction and environment sanitization around tool and provider flows.
\end{itemize}

  \item  \noindent\textit{Systematic risk-avoidance practices} reduce the chance that such a path forms in the first place by limiting risky combinations.
  
\begin{itemize}[leftmargin=*]
  \item \textbf{Scope allowed paths, tools, modes, and data flows.} Permission and policy controls avoid risky chains by making the allowed action set explicit before execution. This practice appears in 100 of 149 secret-to-tool chains, 84 of 128 workspace-to-persistent-state chains, 94 of 137 extension-to-authority chains, 82 of 123 mode-switch-to-privilege chains, 69 of 99 runtime-code chains, 91 of 131 model-to-process chains, 44 of 60 session-to-action chains, and 27 of 45 untrusted-context-to-action chains. For example, \texttt{1Panel-dev/MaxKB} uses token permissions and tool filtering around custom Python tools and generated MCP servers.
  \item \textbf{Make extension and runtime authority reviewable.} MCP/plugin registration and configuration avoid hidden authority chains by making added tools, servers, commands, or generated interfaces visible before they become model-callable. This practice appears in 101 of 137 extension-to-authority chains and 75 of 99 runtime-code chains. For example, \texttt{aaif-goose/goose} exposes MCP extension configuration as a primary control point.
  \item \textbf{Scope browser sessions and domains.} Browser and domain-scoping controls avoid session-to-action chains by limiting which pages, domains, uploads, downloads, storage state, or signed-in sessions can influence later actions. This practice appears in 50 of 60 session-to-action chains and 24 of 45 untrusted-context-to-action chains. For example, \texttt{browser-use/browser-use} includes domain restrictions, sensitive-data filtering, storage-state handling, and upload-path controls.
\end{itemize}

\end{itemize}

\mybox{\textbf{Answer to RQ1.4.} For \textit{risk scenarios} layer, common QA practices include design-time risk reduction and chain interruption at critical points, using mechanisms such as authorization, isolation, and redaction.}

\subsection{{RQ2: Undercovered QA Problems Across the Four RQ1 Layers}}

This RQ examines which QA problems remain undercovered across the four RQ1 layers. We treat a problem as undercovered when repositories expose the problem but lack the most directly relevant QA practice, or when the corresponding practice is implemented only as a local control for a specific route, mode, or artifact type rather than as systematic coverage. Table~\ref{tab:rq2-layer-gaps} summarizes the largest direct gaps by RQ1 dimension.

\begin{table*}[t]
  \centering
  \scriptsize
  \caption{Undercovered QA problems by RQ1 dimension. Percentages show the share of affected projects lacking the corresponding direct QA practice.}
  \label{tab:rq2-layer-gaps}
  \setlength{\tabcolsep}{3pt}
  \begin{tabularx}{\textwidth}{>{\raggedright\arraybackslash}p{0.15\textwidth}>{\raggedright\arraybackslash}p{0.35\textwidth}>{\raggedright\arraybackslash}p{0.26\textwidth}>{\raggedright\arraybackslash}X}
    \toprule
    Dimension & Undercovered QA problems & Projects lacking direct QA practice & Remaining QA gap \\
    \midrule
    RQ1.1 Execution surfaces &
    Credential configuration, network/API routes, file access, command execution, and dynamic-code paths. &
    Secret management/redaction: 36.1\%; permission/policy controls: 34.6\%; sandboxing/isolation: 30.1\% for file access and 30.4\% for command execution. &
    Projects often expose the routes, but do not always define or enforce the boundary that should apply to each route. \\
    RQ1.2 Safeguards &
    Secret handling, file/workspace containment, plugin/MCP extension control, and browser/session scoping. &
    Secret management/redaction: 36.1\%; sandboxing/isolation: 30.1\%; MCP/plugin registration: 27.1\%; browser/domain scoping: 26.9\%. &
    Controls are present, but the direct control for the privileged action is not always present. \\
    RQ1.3 Testing artifacts &
    Adversarial tests for untrusted context, regression or replay tests for multi-step behavior, and boundary tests for high-authority actions. &
    Prompt-injection/adversarial tests: 97.8\%; evaluation/red-team-like paths: 52.4\%; security/safety-oriented tests: 39.7\%. &
    Many tests check components or expected workflows rather than whether model-mediated actions stay inside required boundaries. \\
    RQ1.4 Risk scenarios &
    Credential-flow, command-execution, file-boundary, extension-authority, and runtime-code risk chains. &
    Secret management/redaction: 35.6\%; sandboxing/isolation: 30.5\% for command risk chains and 29.7\% for file-boundary risk chains; MCP/plugin registration: 26.3\%. &
    Risk chains can remain general concerns unless projects turn them into bounded-action requirements and checks. \\
    \bottomrule
  \end{tabularx}
\end{table*}

\textbf{Execution surfaces.} The table shows that the weakest coverage is not for obscure surfaces, but for common routes such as credentials, network/API calls, file access, shell commands, and dynamic code. In these cases, the project exposes an action route but lacks the direct control expected for that route. For example, \texttt{vercel-labs/agent-browser} is counted in the credential gap because its review record includes environment secrets, browser state, injected headers, and cloud profiles reachable from browser/CLI paths, while it does not match the secret-management/redaction practice rule. \texttt{0x4m4/hexstrike-ai} is counted in the file and command gaps because its review record links user/model-controlled paths and shell commands to API/MCP routes, while the corresponding sandbox/isolation practice is not present under our rule. The practical implication is that surface inventory alone is insufficient; each action route needs an attached boundary such as redaction, permission policy, or sandboxing.

\textbf{Safeguards.} The safeguard gaps show that broad controls do not always cover the specific action that needs protection. Secret handling, workspace containment, MCP/plugin extension, and browser session control need different direct practices, but projects often rely on more general approval/authentication or configuration assumptions. For example, \texttt{multica-ai/multica} is counted in the MCP/plugin gap because its review record describes stored per-agent MCP configuration being passed to Claude Code as a local tool channel, but it lacks the direct MCP/plugin registration practice under our reproducible rule. \texttt{0x4m4/hexstrike-ai} is counted in the browser/domain gap because its browser agent can visit arbitrary URLs and return browser-observed secrets, but no browser/domain-scoping control is matched. The remaining QA gap is therefore not simply ``add a safeguard''; it is to bind the right safeguard to the specific sensitive action.

\textbf{Testing artifacts.} The testing gaps are the most direct sign that QA practice has not caught up with agent-specific failure modes. Projects may have ordinary tests, but the missing direct practices are adversarial tests for untrusted context, replay/regression tests for multi-step behavior, and security/safety tests for high-authority actions. For example, \texttt{browser-use/browser-use} is counted in the untrusted-context testing gap because its review record links model-reachable browser JavaScript execution and browser-domain policy defaults to prompt-injection-style risk, while it does not match the prompt-injection/adversarial test rule. \texttt{reworkd/AgentGPT} is similarly counted because external snippets are inserted into summarization prompts, but the matching adversarial test practice is absent. The remaining gap is failure sensitivity: tests should fail when a model-mediated action escapes an approval, sandbox, credential, plugin, file, or browser boundary.

\textbf{Risk scenarios.} The risk-scenario gaps identify chains that need to be converted into QA requirements. These chains include credential flow into tools or subprocesses, commands affecting host state, workspace-boundary crossing, extension-based authority, and runtime-loaded code. For example, \texttt{Fosowl/agenticSeek} is counted in the credential-flow gap because its review record links LLM-generated Python execution to backend secrets, while secret-management/redaction is not matched. \texttt{0x4m4/hexstrike-ai} is counted in the command-execution gap because an unauthenticated API route can reach host shell execution, while sandbox/isolation is not matched. The remaining gap is operationalization: each chain should name the trigger, protected asset, required control, and regression check that would stop the chain.

\textbf{Recommendations.} First, projects should define controls by protected action or asset rather than by interface, so the same secret, file, shell, plugin, browser, and network boundary applies across CLI, API, MCP, browser, headless, and batch routes. Second, each high-authority route or risk chain should have a failure-oriented test or replay case that fails when the required approval, sandbox, policy, redaction, domain scope, or extension check is removed.

\mybox{\textbf{Answer to RQ2.} Many undercovered QA problems are agent-specific rather than traditional software defects: model-mediated tool use, session reuse, extension authority, and untrusted-context propagation often lack corresponding QA practices beyond ordinary tests, access checks, or configuration review.}
\section{Threats}
\label{Threats}

\textbf{Internal validity.} Our analysis is based on repository-level artifacts, including source code, documentation, configuration, tests, and structured review notes. This view captures what developers make inspectable in the repository, but may not include hosted services, provider-side controls, deployment-specific settings, or dependency behavior outside the inspected snapshot. 
{Automated search and LLM-assisted summarization helped identify the candidate artifacts, but may introduce false positives, false negatives, or inconsistent terminology. We mitigate this risk through manual verification of candidate artifacts and a fixed coding schema for quantitative analysis. Because of objective was to inductively develop a repository-level coding framework through iterative open coding, the coding categories evolved during analysis instead of using a fixed predefined codebook. Consequently, we did not compute inter-rater agreement during taxonomy construction. To improve consistency, reviewers continuously refined label definitions, discussed ambiguous cases, and resolved disagreements through consensus before applying the finalized coding rubric across the corpus. This approach follows common qualitative coding practices used for inductive taxonomy development in empirical software engineering.~\cite{diaz2023applying,openja2024empirical}.}

\textbf{Construct validity.} The four analysis layers provide an operational coding rubric rather than a universal taxonomy of agent QA. Some projects could reasonably be grouped differently, especially when one artifact plays multiple roles, such as exposing a tool, constraining its use, and documenting a risk scenario. To reduce this ambiguity, we separated capabilities from QA mechanisms and kept structured records for paths, labels, and review notes. Our counts should be read as repository-visible QA practices, not as measurements of runtime effectiveness.

\textbf{External validity.} The corpus is a fixed May 2026 snapshot of popular open-source GitHub projects and overrepresents coding and browser/computer-use agents. It is not a census of all agent repositories; closed-source, enterprise, and small prototype agents may differ. GitHub stars are descriptive metadata, not a security-maturity measure, and repositories may have changed after collection.

\section{Conclusion}
\label{sec:conclusion}

This paper presented an empirical study of 157 open-source LLM-based agent projects. Using a four-layer coding rubric, we characterized execution surfaces, safeguards, testing artifacts, and risk scenarios from repository artifacts and coded records. The results show that QA practices are common, but their coverage across agent workflows remains uneven. The main gap is not whether projects contain QA artifacts, but whether those artifacts cover the full agent workflow. Future agent QA should make tool authority explicit, connect safeguards and tests to the situations where sensitive actions are reached, and add failure-oriented checks for boundary, adversarial, and multi-step tool-use cases.

\balance
\bibliographystyle{ieeetr}
\bibliography{ref}

@String{Computing = "Computing" }

@String{Computer = "{IEEE} Computer" }

@String{Springer = "Springer-Verlag" }

@article{wang2023surveyagents,
  title={A survey on large language model based autonomous agents},
  author={Wang, Lei and Ma, Chen and Feng, Xueyang and Zhang, Zeyu and Yang, Hao and Zhang, Jingsen and Chen, Zhiyuan and Tang, Jiakai and Chen, Xu and Lin, Yankai and others},
  journal={Frontiers of Computer Science},
  volume={18},
  number={6},
  pages={186345},
  year={2024},
  publisher={Springer}
}

@article{xi2023rise,
  title = {The Rise and Potential of Large Language Model Based Agents: A Survey},
  author = {Xi, Zhiheng and Chen, Wenxiang and Guo, Xin and He, Wei and Ding, Yiwen and Hong, Boyang and Zhang, Ming and Wang, Junzhe and Jin, Senjie and Zhou, Enyu and others},
  journal = {arXiv preprint arXiv:2309.07864},
  year = {2023}
}

@article{yang2024swe,
  title={Swe-agent: Agent-computer interfaces enable automated software engineering},
  author={Yang, John and Jimenez, Carlos and Wettig, Alexander and Lieret, Kilian and Yao, Shunyu and Narasimhan, Karthik and Press, Ofir},
  journal={Advances in Neural Information Processing Systems},
  volume={37},
  pages={50528--50652},
  year={2024}
}

@article{debenedetti2024agentdojo,
  title={Agentdojo: A dynamic environment to evaluate prompt injection attacks and defenses for llm agents},
  author={Debenedetti, Edoardo and Zhang, Jie and Balunovic, Mislav and Beurer-Kellner, Luca and Fischer, Marc and Tram{\`e}r, Florian},
  journal={Advances in Neural Information Processing Systems},
  volume={37},
  pages={82895--82920},
  year={2024}
}

@inproceedings{zhan2024injecagent,
  title={Injecagent: Benchmarking indirect prompt injections in tool-integrated large language model agents},
  author={Zhan, Qiusi and Liang, Zhixiang and Ying, Zifan and Kang, Daniel},
  booktitle={Findings of the Association for Computational Linguistics: ACL 2024},
  pages={10471--10506},
  year={2024}
}

@inproceedings{liu2024agentbench,
  title={Agentbench: Evaluating llms as agents},
  author={Liu, Xiao and Yu, Hao and Zhang, Hanchen and Xu, Yifan and Lei, Xuanyu and Lai, Hanyu and Gu, Yu and Ding, Hangliang and Men, Kaiwen and Yang, Kejuan and others},
  booktitle={International Conference on Learning Representations},
  volume={2024},
  pages={52989--53046},
  year={2024}
}

@article{schick2023toolformer,
  title={Toolformer: Language models can teach themselves to use tools},
  author={Schick, Timo and Dwivedi-Yu, Jane and Dess{\`\i}, Roberto and Raileanu, Roberta and Lomeli, Maria and Hambro, Eric and Zettlemoyer, Luke and Cancedda, Nicola and Scialom, Thomas},
  journal={Advances in neural information processing systems},
  volume={36},
  pages={68539--68551},
  year={2023}
}

@inproceedings{yao2023react,
  title = {ReAct: Synergizing Reasoning and Acting in Language Models},
  author = {Yao, Shunyu and Zhao, Jeffrey and Yu, Dian and Du, Nan and Shafran, Izhak and Narasimhan, Karthik and Cao, Yuan},
  booktitle = {International Conference on Learning Representations},
  year = {2023}
}

@article{karpas2022mrkl,
  title = {MRKL Systems: A Modular, Neuro-Symbolic Architecture that Combines Large Language Models, External Knowledge Sources and Discrete Reasoning},
  author = {Karpas, Ehud and Abend, Omri and Belinkov, Yonatan and Lenz, Barak and Lieber, Opher and Ratner, Nir and Shoham, Yoav and Bata, Hofit and Levine, Yoav and Leyton-Brown, Kevin and Muhlgay, Dor and Rozen, Noam and Schwartz, Erez and Shachaf, Gal and Shalev-Shwartz, Shai and Shashua, Amnon and Tenenholtz, Moshe},
  journal = {arXiv preprint arXiv:2205.00445},
  year = {2022}
}

@inproceedings{shinn2023reflexion,
  title = {Reflexion: Language Agents with Verbal Reinforcement Learning},
  author = {Shinn, Noah and Cassano, Federico and Berman, Edward and Gopinath, Ashwin and Narasimhan, Karthik and Yao, Shunyu},
  booktitle = {Advances in Neural Information Processing Systems},
  year = {2023}
}

@inproceedings{yao2023tree,
  title = {Tree of Thoughts: Deliberate Problem Solving with Large Language Models},
  author = {Yao, Shunyu and Yu, Dian and Zhao, Jeffrey and Shafran, Izhak and Griffiths, Thomas L. and Cao, Yuan and Narasimhan, Karthik},
  booktitle = {Advances in Neural Information Processing Systems},
  year = {2023}
}

@inproceedings{zhou2024lats,
  title = {Language Agent Tree Search Unifies Reasoning Acting and Planning in Language Models},
  author = {Zhou, Andy and Yan, Kai and Shlapentokh-Rothman, Michal and Wang, Haohan and Wang, Yu-Xiong},
  booktitle = {International Conference on Machine Learning},
  year = {2024}
}

@inproceedings{park2023generative,
  title = {Generative Agents: Interactive Simulacra of Human Behavior},
  author = {Park, Joon Sung and O'Brien, Joseph C. and Cai, Carrie J. and Morris, Meredith Ringel and Liang, Percy and Bernstein, Michael S.},
  booktitle = {Proceedings of the 36th Annual ACM Symposium on User Interface Software and Technology},
  year = {2023}
}

@article{
wang2023voyager,
title={Voyager: An Open-Ended Embodied Agent with Large Language Models},
author={Guanzhi Wang and Yuqi Xie and Yunfan Jiang and Ajay Mandlekar and Chaowei Xiao and Yuke Zhu and Linxi Fan and Anima Anandkumar},
journal={Transactions on Machine Learning Research},
issn={2835-8856},
year={2024},
url={https://openreview.net/forum?id=ehfRiF0R3a},
note={}
}

@inproceedings{wu2023autogen,
  title = {AutoGen: Enabling Next-Gen LLM Applications via Multi-Agent Conversation Framework},
  author = {Wu, Qingyun and Bansal, Gagan and Zhang, Jieyu and Wu, Yiran and Li, Beibin and Zhu, Erkang and Jiang, Li and Zhang, Xiaoyun and Zhang, Shaokun and Liu, Jiale and others},
  booktitle = {Conference on Language Modeling},
  year = {2024}
}

@inproceedings{hong2023metagpt,
      title={Meta{GPT}: Meta Programming for A Multi-Agent Collaborative Framework},
      author={Sirui Hong and Mingchen Zhuge and Jonathan Chen and Xiawu Zheng and Yuheng Cheng and Jinlin Wang and Ceyao Zhang and Zili Wang and Steven Ka Shing Yau and Zijuan Lin and Liyang Zhou and Chenyu Ran and Lingfeng Xiao and Chenglin Wu and J{\"u}rgen Schmidhuber},
      booktitle={The Twelfth International Conference on Learning Representations},
      year={2024},
      url={https://openreview.net/forum?id=VtmBAGCN7o}
}

@inproceedings{qian2023chatdev,
    title = "{C}hat{D}ev: Communicative Agents for Software Development",
    author = "Qian, Chen  and
      Liu, Wei  and
      Liu, Hongzhang  and
      Chen, Nuo  and
      Dang, Yufan  and
      Li, Jiahao  and
      Yang, Cheng  and
      Chen, Weize  and
      Su, Yusheng  and
      Cong, Xin  and
      Xu, Juyuan  and
      Li, Dahai  and
      Liu, Zhiyuan  and
      Sun, Maosong",
    editor = "Ku, Lun-Wei  and
      Martins, Andre  and
      Srikumar, Vivek",
    booktitle = "Proceedings of the 62nd Annual Meeting of the Association for Computational Linguistics (Volume 1: Long Papers)",
    month = aug,
    year = "2024",
    address = "Bangkok, Thailand",
    publisher = "Association for Computational Linguistics",
    url = "https://aclanthology.org/2024.acl-long.810/",
    doi = "10.18653/v1/2024.acl-long.810",
    pages = "15174--15186",
}

@inproceedings{greshake2023indirect,
author = {Greshake, Kai and Abdelnabi, Sahar and Mishra, Shailesh and Endres, Christoph and Holz, Thorsten and Fritz, Mario},
title = {Not What You've Signed Up For: Compromising Real-World LLM-Integrated Applications with Indirect Prompt Injection},
year = {2023},
isbn = {9798400702600},
publisher = {Association for Computing Machinery},
address = {New York, NY, USA},
url = {https://doi.org/10.1145/3605764.3623985},
doi = {10.1145/3605764.3623985},
booktitle = {Proceedings of the 16th ACM Workshop on Artificial Intelligence and Security},
pages = {79–90},
numpages = {12},
location = {Copenhagen, Denmark},
series = {AISec '23}
}

@inproceedings{liu2024formalizing,
  title = {Formalizing and Benchmarking Prompt Injection Attacks and Defenses},
  author = {Liu, Yupei and Jia, Yuqi and Geng, Runpeng and Jia, Jinyuan and Gong, Neil Zhenqiang},
  booktitle = {USENIX Security Symposium},
  year = {2024}
}

@article{liu2023houyi,
  title = {Prompt Injection Attack against LLM-Integrated Applications},
  author = {Liu, Yi and Deng, Gelei and Li, Yuekang and Wang, Kailong and Wang, Zihao and Wang, Xiaofeng and Zhang, Tianwei and Liu, Yepang and Wang, Haoyu and Zheng, Yan and Zhang, Leo Yu and Liu, Yang},
  journal = {arXiv preprint arXiv:2306.05499},
  year = {2023}
}

@inproceedings{ruan2023toolemu,
  title = {Identifying the Risks of LM Agents with an LM-Emulated Sandbox},
  author = {Ruan, Yangjun and Dong, Honghua and Wang, Andrew and Pitis, Silviu and Zhou, Yongchao and Ba, Jimmy and Dubois, Yann and Maddison, Chris J. and Hashimoto, Tatsunori},
  booktitle = {International Conference on Learning Representations},
  year = {2024}
}

@inproceedings{zhang2025asb,
  title = {Agent Security Bench (ASB): Formalizing and Benchmarking Attacks and Defenses in LLM-based Agents},
  author = {Zhang, Hanrong and Huang, Jingyuan and Mei, Kai and Yao, Yifei and Wang, Zhenting and Zhan, Chenlu and Wang, Hongwei and Zhang, Yongfeng},
  booktitle = {International Conference on Learning Representations},
  year = {2025}
}

@article{hasan2025mcpfirstglance,
  title = {Model Context Protocol (MCP) at First Glance: Studying the Security and Maintainability of MCP Servers},
  author = {Hasan, Mohammed Mehedi and Li, Hao and Fallahzadeh, Emad and Rajbahadur, Gopi Krishnan and Adams, Bram and Hassan, Ahmed E.},
  journal = {arXiv preprint arXiv:2506.13538},
  year = {2025}
}

@article{yang2025mcpsecbench,
  title = {MCPSecBench: A Systematic Security Benchmark and Playground for Testing Model Context Protocols},
  author = {Yang, Yixuan and Gao, Cuifeng and Wu, Daoyuan and Chen, Yufan and Li, Yingjiu and Wang, Shuai},
  journal = {arXiv preprint arXiv:2508.13220},
  year = {2025}
}

@article{debenedetti2025camel,
  title = {Defeating Prompt Injections by Design},
  author = {Debenedetti, Edoardo and Shumailov, Ilia and Fan, Tianqi and Hayes, Jamie and Carlini, Nicholas and Fabian, Daniel and Kern, Christoph and Shi, Chongyang and Terzis, Andreas and Tram{\`e}r, Florian},
  journal = {arXiv preprint arXiv:2503.18813},
  year = {2025}
}

@inproceedings{li2025dissonances,
	author = {Zichuan Li and Jian Cui and Xiaojing Liao and Luyi Xing},
	title = {Les Dissonances: Cross-Tool Harvesting and Polluting in Pool-of-Tools Empowered LLM Agents},
	booktitle = {33nd Annual Network and Distributed System Security Symposium, {NDSS}
	2026, San Diego, California, USA, February 24-27, 2026},
	year = {2026}, 
	month = {February},
	address = {San Diego, CA}
}

@misc{owasp2025llm,
  title = {OWASP Top 10 for Large Language Model Applications},
  author = {{OWASP Foundation}},
  year = {2025},
  howpublished = {\url{https://owasp.org/www-project-top-10-for-large-language-model-applications/}}
}

@article{aloraini2019sast,
  title = {An Empirical Study of Security Warnings from Static Application Security Testing Tools},
  author = {Aloraini, Bushra and Nagappan, Meiyappan and German, Daniel M. and Hayashi, Shinpei and Higo, Yoshiki},
  journal = {Journal of Systems and Software},
  volume = {158},
  pages = {110427},
  year = {2019},
  doi = {10.1016/j.jss.2019.110427}
}

@inproceedings{zahan2022weaklinks,
  title = {What are Weak Links in the npm Supply Chain?},
  author = {Zahan, Nusrat and Zimmermann, Thomas and Godefroid, Patrice and Murphy, Brendan and Maddila, Chandra Shekhar and Williams, Laurie A.},
  booktitle = {Proceedings of the 44th International Conference on Software Engineering: Software Engineering in Practice},
  pages = {331--340},
  year = {2022},
  doi = {10.1109/ICSE-SEIP55303.2022.9794068}
}

@article{hasan2025testingpractices,
  title={An empirical study of testing practices in open source AI agent frameworks and agentic applications},
  author={Hasan, Mohammed Mehedi and Li, Hao and Fallahzadeh, Emad and Rajbahadur, Gopi Krishnan and Adams, Bram and Hassan, Ahmed E},
  journal={Empirical Software Engineering},
  volume={31},
  number={5},
  pages={124},
  year={2026},
  publisher={Springer}
}

@inproceedings{li2023camel,
author = {Li, Guohao and Al Kader Hammoud, Hasan Abed and Itani, Hani and Khizbullin, Dmitrii and Ghanem, Bernard},
title = {CAMEL: communicative agents for "mind" exploration of large language model society},
year = {2023},
publisher = {Curran Associates Inc.},
address = {Red Hook, NY, USA},
booktitle = {Proceedings of the 37th International Conference on Neural Information Processing Systems},
articleno = {2264},
numpages = {18},
location = {New Orleans, LA, USA},
series = {NIPS '23}
}

@article{seaman1999qualitative,
  title = {Qualitative Methods in Empirical Studies of Software Engineering},
  author = {Seaman, Carolyn B.},
  journal = {IEEE Transactions on Software Engineering},
  volume = {25},
  number = {4},
  pages = {557--572},
  year = {1999},
  doi = {10.1109/32.799955}
}

@inproceedings{cruzes2011thematic,
  title = {Recommended Steps for Thematic Synthesis in Software Engineering},
  author = {Cruzes, Daniela S. and Dyb{\aa}, Tore},
  booktitle = {Proceedings of the 2011 International Symposium on Empirical Software Engineering and Measurement},
  pages = {275--284},
  year = {2011},
  doi = {10.1109/ESEM.2011.36}
}

@article{diaz2023applying,
  title={Applying inter-rater reliability and agreement in collaborative grounded theory studies in software engineering},
  author={D{\'\i}az, Jessica and P{\'e}rez, Jorge and Gallardo, Carolina and Gonz{\'a}lez-Prieto, {\'A}ngel},
  journal={Journal of Systems and Software},
  volume={195},
  pages={111520},
  year={2023},
  publisher={Elsevier}
}

@article{openja2024empirical,
  title={An empirical study of testing machine learning in the wild},
  author={Openja, Moses and Khomh, Foutse and Foundjem, Armstrong and Jiang, Zhen Ming and Abidi, Mouna and Hassan, Ahmed E},
  journal={ACM transactions on software engineering and methodology},
  volume={34},
  number={1},
  pages={1--63},
  year={2024},
  publisher={ACM New York, NY}
}

\end{document}